# Anomalous and Spin Hall Effects in a Magnetic Tunnel Junction with Rashba Spin-Orbit Coupling

A.V. Vedyayev[1], M.S. Titova[1], N.V. Ryzhanova[1], M.Ye. Zhuravlev[2,3,4], and E.Y. Tsymbal[3]

[1]*Department of Physics, M. V. Lomonosov Moscow State University, 119899 Moscow, Russia*

[2]*Kurnakov Institute for General and Inorganic Chemistry, Russian Academy of Sciences, 119991 Moscow, Russia*

[3]*Department of Physics and Astronomy & Nebraska Center for Materials and Nanoscience, University of Nebraska, Lincoln, Nebraska 68588, USA*

[4]*Faculty of Liberal Arts and Sciences, St. Petersburg State University,190000 St. Petersburg, Russia*

Anomalous and spin Hall effects are investigated theoretically for a magnetic tunnel junction where the applied voltage produces a Rashba spin-orbit coupling within the tunneling barrier layer. The ferromagnetic electrodes are the source of the spin-polarized current. The tunneling electrons experience a spin-orbit coupling inside the barrier due to the applied electrical field. Charge and spin Hall currents are calculated as functions of the position inside the barrier and the angle between the magnetizations of the electrodes. We find that both charge and spin Hall currents are located inside the barrier near the interfaces. The dependence of the currents on magnetic configuration of the magnetic tunnel junction makes possible the manipulation by the Hall currents via rotation of the magnetization of the electrodes.

### I. Introduction

Spin-orbit interaction (SOI) allows the manipulation of spin-polarized currents in thin-film heterostructures [1]. Such heterostructures attract great interest due to potential applications in spintronics [2]. A number of spintronic devices have been proposed which rely on the SOI. Among them are spin field-effect-transistors [3], spin-interference devices [4], and spin filters [5]. SOI is responsible for many physical properties and phenomena which are subjects of intensive investigations, including the topological insulators, [6] the Anomalous Hall effect (AHE) [7], the Spin-Hall effect (SHE) [8], electrically controlled interface magnetocrystalline anisotropy [9], and tunneling anisotropic magnetoresistance [10 – 12].

There are several mechanisms of SOI in layered structures. The Dresselhaus SOI [13] originates from the broken inversion symmetry of a crystal. The Rashba SOI [14] is proportional to the gradient of electrical potential which can be either intrinsic or extrinsic. The intrinsic Rashba effect is well known for surfaces and interfaces [15 – 20] and originates from the electric potential gradient across the interface (surface). The extrinsic contribution allows tuning the Rashba spin-orbit coupling by the applied voltage [21, 22]. Therefore, the Rashba SOI may serve as an important control parameter for the devices based on manipulation of the electron spin.

So far, the AHE and SHE have been studied in metallic or semiconducting systems where the current flows in the plane of the layers. In this work we analyze a different situation. We consider a magnetic tunnel junction (MTJ) where the voltage drop takes place across a tunneling barrier and the electric current flows perpendicular to the plane. The Rashba SOI is produced by the electric field in the insulating barrier layer. Due to the Rashba SOI, the tunneling electrons with the opposite spin projection scatter in the opposite directions. Therefore, charge and spin Hall currents are generated in the direction perpendicular to the driving electric field in the barrier due to the spin-polarized current across the MTJ. We find that, due to the evanescent states being responsible for the tunneling current, the amplitudes of the charge and spin Hall currents are largest near metal/insulator interfaces and decay exponentially into the barrier. The magnitude and direction of the charge and spin Hall currents can be controlled by the orientation of the electrode's magnetizations. A related effect was investigated in ref. [23], where spin-orbit scattering was assumed to be caused by the electron scattering on impurities inside the barrier in a MTJ.

## II. A model for MTJ

We consider a MTJ which consists of two identical semi-infinite ferromagnetic electrodes separated by an insulating barrier layer, as shown in Fig. 1. The MTJ is assumed to be infinite in *x-z* plane. The electronic structure of the MTJ is described using a free-electron model with the exchange splitting of the conducting band. We assume that direct quantum-mechanical tunneling of electrons is the dominant transport mechanism in the system. To take into account the SOI in the barrier, we adopt formal approach which has been previously applied in refs. [24, 25] to describe AHE in the disordered ferromagnets. The Hamiltonians describing the ferromagnetic electrodes ($H^F$) and the barrier ($H^B$) are as follows:

$$H^F_{\sigma\sigma'} = -\frac{\hbar^2}{2m}\Delta\delta_{\sigma\sigma'} + E_c^{L,R} - J_{exch}\left(\cos\theta \cdot \tau^z_{\sigma\sigma'} + \sin\theta \cdot \tau^x_{\sigma\sigma'}\right)$$

$$H^B_{\sigma\sigma'} = -\frac{\hbar^2}{2m}\Delta\delta_{\sigma\sigma'} + U(y) - i\lambda\left(\vec{\tau}_{\sigma\sigma'}(\nabla V \times \nabla)\right)$$

(1)

where $\lambda$ is the SOI constant, $\hbar$ is the Planck constant, $m$ is the electron mass, $\vec{\tau}$ are the Pauli matrices, $J_{exch}$ is the exchange splitting of the conducting band, $E_c^L = 0, E_c^R = -eV$ determine the position of the conducting band in the left ($L$) and right ($R$) electrodes. We approximate the electric potential by a constant in the electrodes and by a linear function inside the barrier. This electric potential is overlapped with a rectangular tunneling potential of height $U_0$ so that the total potential takes a trapezoidal form, $U(y) = U_0 - eEy$. The applied bias is described by the constant electric field $E$ inside the insulating spacer. The SOI term is proportional to the gradient of the electric potential $V$ inside the barrier. We assume that the magnetizations of the ferromagnetic electrodes lie in the plane of the MTJ and use angle $\theta$ to describe the relative orientation of the magnetizations $\vec{M}_L, \vec{M}_R$ in the left and right electrodes. Both vectors $\vec{M}_L, \vec{M}_R$ are parallel to the interfaces. We set $\theta = 0$ for the left electrode and arbitrary value of $\theta$ for the right electrode. The SOI constant $\lambda$ in the simplest case equals to [1]

$$\lambda = \frac{\hbar^2}{(mc)^2}\frac{1}{a^2}$$

(2)

where $a$ is the lattice constant and $c$ is the speed of light.

To calculate the current we need to find the wave function $\Psi$ of the system. The exact wave functions in the presence of SOI can be calculated according to refs. [26, 27]. However, in our calculations we adopt a simpler approach. We consider the SOI as a perturbation, and use the wave function in the barrier without SOI for the calculation of charge and spin currents. The wave function of the system can be found exactly, as well as in the WKB approximation [28]. Here, for simplicity, we employ the latter approach.

### III. Calculation of charge and spin Hall currents

Due to the Rashba SOI in the barrier, electrons acquire velocity components perpendicular to the direction of the driving field. Without spin-orbit coupling the current would flow strictly in the *y*-direction (in our notation of the coordinate system, Fig. 1). To calculate the charge current we use the quantum-mechanical expression for the velocity operator as follows:

$$\vec{v} = -\frac{i}{\hbar}[\vec{r}, H^B] =$$
$$= \frac{-i\hbar}{m}\left\{\frac{\partial}{\partial x}, \frac{\partial}{\partial y}, \frac{\partial}{\partial z}\right\}\hat{I} + \frac{\lambda}{\hbar}eE\{\tau_z, 0, -\tau_x\} \quad (3)$$

where $\hat{I}$ is two-by-two identity matrix in the spin space, $E = |\vec{\nabla}V|$ is the electric field inside the barrier directed along the *y*-direction, $e$ is the electron charge. To calculate the Hall current, we keep only the second term in Eq. (3).

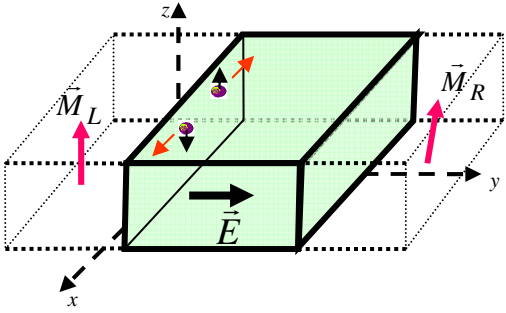

**Fig. 1**. Schematic illustration of a magnetic tunnel junction. $\vec{M}_L, \vec{M}_R$ are the (non-collinear) magnetizations of the left and right ferromagnetic electrodes.

The total charge current is given by

$$J_e = J_L^\uparrow + J_L^\downarrow + J_R^\uparrow + J_R^\downarrow, \quad (4)$$

where indices $L, R$ specify the boundary conditions for the wave function, to distinguish the contributions to the current from the electrons with spin $\sigma = \uparrow$ or $\downarrow$ arriving from the left and right electrodes. Since we consider the non-collinear magnetization of the MTJ, both components of the wave function are non-zero. The partial $\alpha$- component ($\alpha = x, z$) of the current is calculated by the summation over all filled states. The charge Hall current originating from electrons coming from the left electrode is as follows:

$$\left(J_L^\sigma\right)_\alpha(x,\theta) = \frac{\lambda e^2 m}{(2\pi\hbar)^2} \times$$
$$\iint q\,dq\left(\Psi_{L,\sigma}(\vec{v}_\alpha \Psi_{L,\sigma})^* + \Psi_{L,\sigma}^*(\vec{v}_\alpha \Psi_{L,\sigma})\right)f(\varepsilon)d\varepsilon \quad (5)$$

where $f(\varepsilon)=1/(1+\exp[(\varepsilon-E_F)/T])$ is the Fermi distribution function, $E_F$ is the Fermi energy, $T$ is temperature, and the wave function $\Psi_{L,\sigma}$ is a two-component spinor. The integration in Eq. (5) is performed over the electron energy and the in-plane momentum $\vec{q}$. The wave function in Eq. (5) is taken in "q-y" representation so that it depends on coordinate $y$. Also, the wave function depends on the mutual orientation of the ferromagnetic electrodes' magnetizations. The current due to the electrons incoming from the right electrode, $J_R^\sigma$, is calculated using the same formula as Eq. (5) with the replacement $\Psi_{L,\sigma} \to \Psi_{R,\sigma}$ and $f(\varepsilon) \to f(\varepsilon-eV)$.

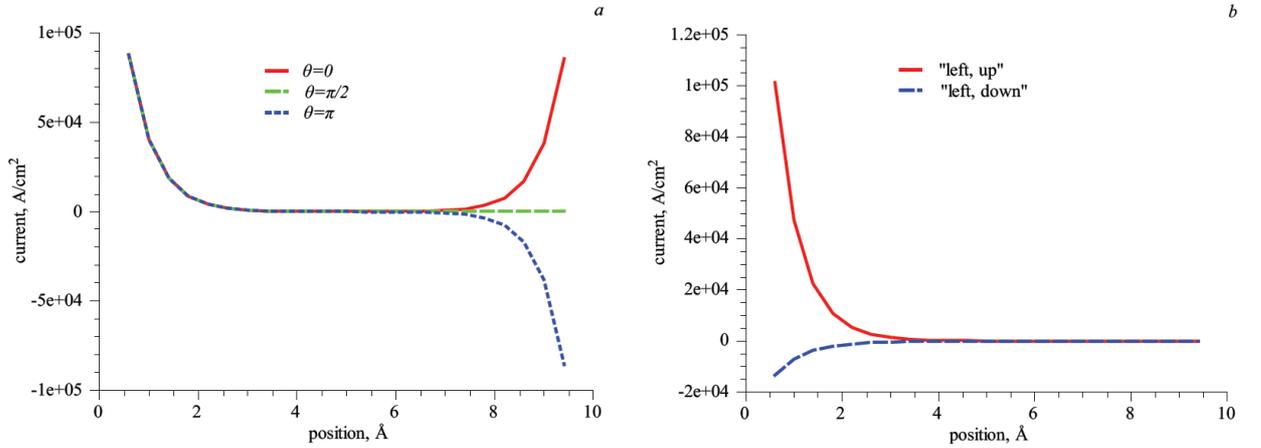

**Fig. 2.** Charge Hall current $J_x$ as a function of the position inside the barrier (a) for different angles $\theta$ and partial contributions to the Hall current from the "spin-up" and "spin-down" electrons coming from the left (b). The calculations are performed using the following parameters: barrier thickness $d=10$ Å, the Fermi energy $E_F=2.0$ eV, the exchange splitting $J_{exch}=1.0$ eV, the applied bias $V=1.0$ V, the barrier height $U=2.0$ eV.

We calculate the charge Hall current as a function of the position inside the barrier and the angle $\theta$ between the magnetization of the electrodes. The calculations are performed using representative parameters typical for experiments: barrier thickness $d=10$ Å, the applied voltage $V=1.0$ V and the barrier height $U=2.0$ eV. Fig. 2a displays the dependence of charge Hall current on $y$, $0<y<d$, for three different values of $\theta$. Our calculations reveal that the charge Hall current is non-zero inside the barrier. This is due to the evanescent states that provide a non-zero contribution to the current. As seen from Fig. 2a, the magnitude the charge Hall current is much higher near the interfaces and decays exponentially into the barrier region. We also see that the sign of the current near the right interface depends on the orientation of the right electrode magnetization. Fig. 2b displays the partial contributions of the currents corresponding to the waves coming from the left with "up" and "down" z-projection of the spin. As expected, the dominant contribution to the charge Hall current near the left interface comes from the region near the left interface and the contributions from the electrons with opposite spin orientations have an opposite sign with the majority-spin ("up") electron contribution dominating.

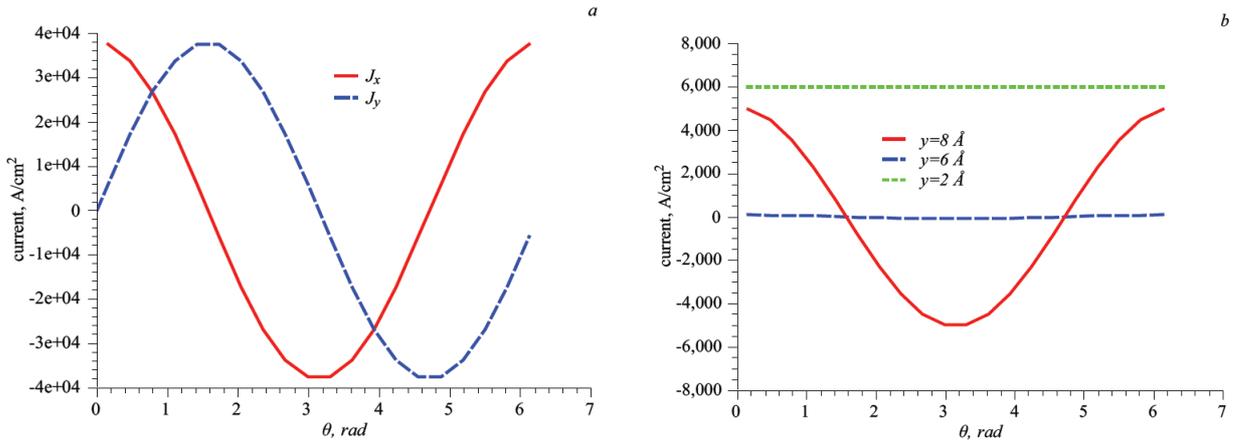

**Fig. 3.** The charge Hall current as a function of angle $\theta$ between magnetization of the electrodes for different positions inside the barrier. (a) $J_x$ (solid line) and $J_z$ (dashed line) for $y = 9$ Å (1 Å from the right interface) are shown; (b) $J_x$ for $y = 8$ Å (solid line), $y = 6$ Å (dashed line) and $y = 2$ Å (dotted line).

Fig. 3 shows the $\theta$ dependence of the *x*- and *z*- components of the charge Hall current for a fixed position inside the barrier. The total Hall current near the right interface is perpendicular to the magnetization of the right electrode. The direction of the Hall current rotates with the electrode's magnetization. Although the magnitude of the current strongly depends on the position in the barrier, all the curves follow the $\cos\theta$ dependence. This angular dependence resembles that of the resistance of the magnetic valve [29, 30]. However, due to the charge Hall current in the barrier being a sum of the currents from the two electrodes (see Eq. (4)) and each contribution decreasing exponentially with the distance from the respective electrode, the local angular dependence is largely determined by the contribution from the electrode which is closer. Therefore, the observed $\cos\theta$ dependence does not reflect the spin valve effect, but rather the magnetization projection in the closest electrode. Correspondingly, the *z*-component of the Hall current near the right interface obeys $\sin\theta$ dependence, as is seen from Fig. 3a (dashed curve). This is also evident from the fact that at positions close to the right electrode (*y* = 9Å, Fig. 3*a*, and *y* = 8Å and 6Å, Fig. 3*b*) the direction (sign) of the charge Hall current changes when the right electrode magnetization is rotated, whereas close to the left electrode (*y* = 2Å, Fig. 3b) the sign does not change, reflecting the dominant contribution to the current from the left electrode whose magnetization orientation is assumed to be fixed.

Next, we calculate the spin Hall current according to the standard approach (see, e.g., ref. [31]). The spin current is a tensor, $J_{\beta\alpha}$, where the first index refers to the spin component and the second index refers to the electron velocity component. As in case of the charge Hall current, we hold only the second term in the expression for the velocity (3) when calculate the spin Hall current. The total spin current consists of four contributions which are distinguished by the boundary conditions:

$$J_{\beta\alpha} = (J_L^{\uparrow})_{\beta\alpha} + (J_L^{\downarrow})_{\beta\alpha} + (J_R^{\uparrow})_{\beta\alpha} + (J_R^{\downarrow})_{\beta\alpha};$$

$$\left(J_L^{\sigma}\right)_{\beta\alpha}(x,\theta) = \frac{\lambda m}{(2\pi\hbar)^2} \times$$

$$\iint q dq \, \mathrm{Re}\left(\Psi_{L,\sigma} \tau_\beta \vec{v}_\alpha \Psi_{L,\sigma}^*\right) f(\varepsilon) d\varepsilon +$$

$$\iint q dq \, \mathrm{Re}\left(\Psi_{R,\sigma} \tau_\beta \vec{v}_\alpha \Psi_{R,\sigma}^*\right) f(\varepsilon - eV) d\varepsilon$$

(6)

The *y*-component of the second term in Eq. 3 for the velocity operator is equal to zero, and hence there are six components of the spin current tensor which can be non-vanishing, namely $J_{zx} = -J_{xz}$, $J_{yx}$, $J_{yz}$, $J_{xx} = J_{zz}$. In fact, in our geometry, when the current flows in the *y*-direction and the magnetization is constrained to the *x-z* plane, only $J_{zx} = -J_{xz}$ and $J_{yz}$ are non-zero. Fig. 4 shows the calculated $J_{zx} = -J_{xz}$ and $J_{yz}$ components of the spin current as a function of position in the barrier for $\theta = 0$, $\theta = \pi/2$ and $\theta = \pi$. It is seen from Fig. 4a that $J_{zx} = -J_{xz}$ does not depend on the angle of the magnetization in the electrode (in Fig. 4a all the three curves coincide). Normally, these components are referred to as the "spin Hall current". In our case, we find that the $J_{yz}$ component of the spin current tensor is also non-zero. As seen from Fig. 4b, it is both position and angle dependent, and resembles the corresponding dependences for the charge current (Fig. 2).

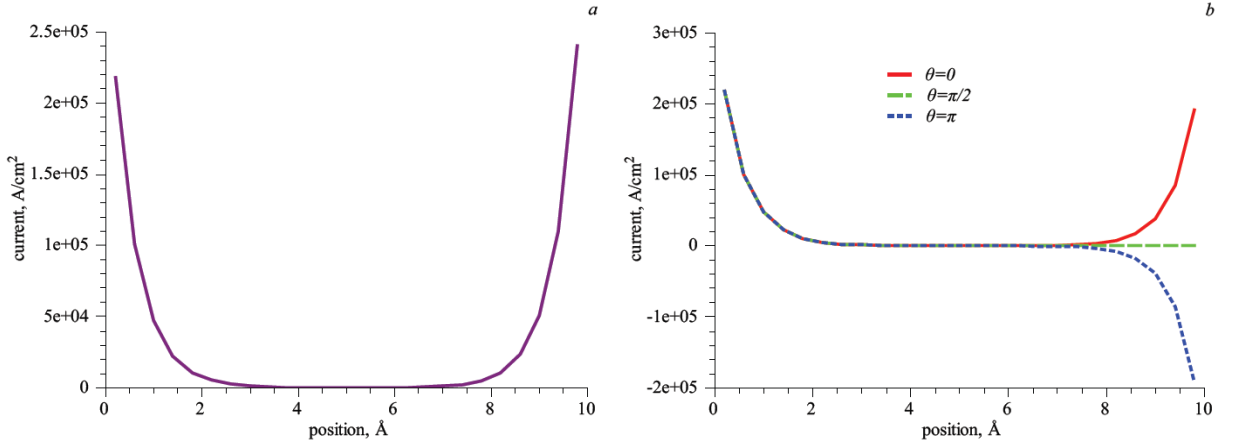

**Fig 4:** $J_{zx} = -J_{xz}$ (*a*) and $J_{yz}$ (*b*) components of the spin Hall current as a function of position in the barrier for different angles $\theta$.

We note that the anomalous charge and spin Hall currents at the metal/insulator interfaces inside the metal have been investigated previously, in particularly, in ref. [32]. Those currents are due to the SOI produced by the intrinsic fields at the metal/insulator interfaces which can be very high. Yet, in ref. [32] the charge and spin currents associated with these fields flow along the interfaces in the metal layers.

## IV. Conclusions

We have investigated theoretically the anomalous and spin Hall effects for a magnetic tunnel junction under non-zero bias. The charge and spin Hall currents originate from the Rashba spin-orbit coupling of the tunneling electrons due to an applied electrical field. The distribution of the charge and spin Hall currents across the barrier were calculated. We found that the currents are largest at the barrier/ferromagnet interfaces and decay exponentially with distance from the interfaces, following the evanescent character of the states in the barrier. The directions of the charge Hall current and the $J_{yz}$ component of the spin current near the interface depend on the magnetization orientation in the adjacent ferromagnetic electrode. Therefore, one can control the charge and spin Hall currents in the barrier via the electrodes' magnetization rotation. Finally, it is worth noting that the current density flowing through the barrier decreases exponentially as the barrier width increases, whereas the density of the charge and spin Hall currents considered here decreases as the inverse thickness of the barrier. Therefore, the density of the Hall current can exceed the density of the current through the barrier.

**Acknowledgements**

This research was supported by the Russian Foundation for Fundamental Research via grant 13-02-01452. Research at University of Nebraska was supported by the National Science Foundation through the Experimental Program to Stimulate Competitive Research (Grant No. EPS-1010674).